\documentclass[sigconf]{acmart} 
\AtBeginDocument{%
  }

\usepackage{xcolor}
\usepackage{booktabs}
\usepackage{graphicx}
\usepackage{multirow}
\usepackage{siunitx}

\definecolor{pastelBlue}{RGB}{200, 230, 240}
\definecolor{pastelGreen}{RGB}{200, 245, 200}
\definecolor{pastelYellow}{RGB}{255, 245, 200}
\definecolor{pastelPink}{RGB}{255, 210, 220}
\definecolor{pastelPurple}{RGB}{235, 220, 240}
\definecolor{pastelOrange}{RGB}{255, 225, 200}
\definecolor{pastelMint}{RGB}{210, 250, 225}
\definecolor{pastelGray}{RGB}{235, 235, 235}
\definecolor{pastelTeal}{RGB}{210, 245, 245}
\definecolor{pastelCoral}{RGB}{255, 200, 185}
\definecolor{darkBlue}{RGB}{30, 70, 120}
\definecolor{darkGreen}{RGB}{20, 100, 50}
\definecolor{darkYellow}{RGB}{180, 140, 20}
\definecolor{darkPink}{RGB}{170, 40, 80}
\definecolor{darkPurple}{RGB}{90, 40, 110}
\definecolor{darkOrange}{RGB}{180, 80, 20}
\definecolor{darkMint}{RGB}{0, 120, 90}
\definecolor{darkGray}{RGB}{80, 80, 80}
\definecolor{darkTeal}{RGB}{0, 100, 110}
\definecolor{darkCoral}{RGB}{190, 70, 50}
\definecolor{green1}{HTML}{A8E6CF}
\definecolor{green2}{HTML}{D4F5C7}
\definecolor{red1}{HTML}{E8A1A1}
\definecolor{red2}{HTML}{F5BDBD}
\definecolor{yellow1}{HTML}{FFF7C2} 
\definecolor{orange1}{HTML}{FFE4B2}
\definecolor{blue1}{HTML}{C9E4F6}
\definecolor{purple1}{HTML}{D9C9EB}
\definecolor{pink1}{HTML}{F4C6D7}
\definecolor{cgreen}{HTML}{228B22}

\copyrightyear{2026}
\acmYear{2026}
\setcopyright{cc}
\setcctype{by}
\acmConference[CHI EA '26]{Extended Abstracts of the 2026 CHI Conference on Human Factors in Computing Systems}{April 13--17, 2026}{Barcelona, Spain}
\acmBooktitle{Extended Abstracts of the 2026 CHI Conference on Human Factors in Computing Systems (CHI EA '26), April 13--17, 2026, Barcelona, Spain}
\acmDOI{10.1145/3772363.3798677}
\acmISBN{979-8-4007-2281-3/2026/04}
\begin{document}
\renewcommand\footnotetextcopyrightpermission[1]{}


\title{From Seeing it to Experiencing it: Interactive Evaluation of Intersectional Voice Bias in Human–AI Speech Interaction}


\author{Shree Harsha Bokkahalli Satish}
\affiliation{%
  \institution{KTH Royal Institute of Technology}
  \city{Stockholm}
  \country{Sweden}}
\email{shbs@kth.se}

\author{Maria Teleki}
\affiliation{%
  \institution{Texas A\&M University}
  \city{College Station}
  \country{USA}}
\email{mariateleki@tamu.edu}

\author{Christoph Minixhofer}
\affiliation{%
  \institution{University of Edinburgh}
  \city{Edinburgh}
  \country{UK}
}
\email{christoph.minixhofer@ed.ac.uk}

\author{Ondrej Klejch}

\affiliation{%
  \institution{University of Edinburgh}
  \city{Edinburgh}
  \country{UK}
}
\email{o.klejch@ed.ac.uk}

\author{Peter Bell}

\affiliation{%
  \institution{University of Edinburgh}
  \city{Edinburgh}
  \country{UK}
}

\email{peter.bell@ed.ac.uk}

\author{Éva Székely}
\affiliation{%
  \institution{KTH Royal Institute of Technology}
  \city{Stockholm}
  \country{Sweden}}
\email{szekely@kth.se}
\renewcommand{\shortauthors}{Bokkahalli Satish \textit{et al.}}

\begin{abstract}
SpeechLLMs process spoken language directly from audio, but accent and vocal identity cues can lead to biased behaviour. Current bias evaluations often miss how such bias manifests in end-to-end speech interactions and how users experience it. We distinguish quality-of-service disparities (e.g., off-topic or low-effort responses) from content-level bias in coherent outputs, and examine intersectional effects of accent and perceived gender. In this work, we explore a two-part evaluation approach: (1) a controlled test cohort spanning six accents and two gender presentations, analysed with judge-free prompt–response metrics, and (2) an interactive study design using voice conversion to let users experience identical content through different vocal identities. Across two studies (Interactive, N=24; Observational, N=19), we find that voice conversion increases trust and acceptability for benign responses and encourages perspective-taking, while automated analysis in search of quality-of-service disparities, reveals \textit{accent×gender} disparities in alignment and verbosity across SpeechLLMs. These results highlight voice conversion for probing and experiencing intersectional voice bias while our evaluation suite provides richer bias evaluations for spoken conversational AI.
\end{abstract}




\maketitle

\section{Introduction}

Speech-based interfaces are central to a new generation of conversational systems (e.g., voice assistants, smart speakers, smart glasses) \cite{chatgpt_voice_features, google_gemini_assistant, toyota_intelligent_assistant_2022, google_home_welcome_2025, amazon_echo_category, meta_ai_glasses_2025}. Speech Large Language Models (SpeechLLMs) process spoken language directly from audio (rather than a cascaded ASR $\rightarrow$ LLM $\rightarrow$ TTS pipeline)~\cite{cui_recent_2025}, enabling sensitivity to prosody and other paralinguistic cues -- but also exposing models to accent and demographic signals that can shape model behaviour.

In this paper, we study \textit{intersectional voice bias} in SpeechLLMs through a \textbf{two-part evaluation approach} that combines (1) \textbf{controlled, judge-free testing} across a cohort of accents and gender presentations with prompt--response metrics, and (2) \textbf{interactive, user-facing evaluation} using \textbf{voice conversion (VC)} to let participants experience identical linguistic content through different vocal identities. We frame bias in speech interaction along two levels. \textbf{Interaction Failure / QoS bias} refers to disparities in whether the system produces a usable interaction at all (e.g., off-topic, low-effort, misaligned, incoherent, or prematurely terminated responses). \textbf{Content-level bias} refers to harmful or differential \emph{content} when outputs are otherwise coherent (e.g., stereotyping, different recommendations/advice, or differential politeness). This distinction clarifies which harms can be detected via quality-of-service (QoS) metrics versus which require analysis of semantic content and user interpretation.

Previous work on identity and voice has often focused on disparities in automatic speech recognition across demographic groups \cite{koenecke2020racial, Metz2020, mengesha2021don, feng2024towards, harris-etal-2024-modeling, cunningham2025toward, choi2025fairnessautomaticspeechrecognition}, while newer work increasingly considers voice variation and bias in SpeechLLMs~\cite{bokkahalli2025bias, lameris2025lost, wu2025evaluating, tam2025medvoicebias}. To isolate vocal identity effects while holding lexical content constant, we use \textbf{Voice Conversion (VC)} to alter accent and perceived gender while preserving linguistic content~\cite{sisman2020overview, liu2024zero}.

In our own prior work, we introduced \emph{Hear Me Out}~\cite{HearMeOut2025}, an interactive system that used voice conversion to let users experience SpeechLLM interactions through different vocal identities. Presented as a Show \& Tell demonstration, this system enabled informal, exploratory interaction with AI responses to voice-converted user queries, highlighting the potential of voice conversion as a means for surfacing speech identity-based bias. 
This paper presents the first empirical study of this voice-conversion based evaluation approach. In particular, we test whether users are more accepting of potentially concerning AI advice as a response to their own converted voice versus observing the same concerning responses in a third-person perspective. We ask:
\begin{itemize}
\item (\textbf{RQ1}) How does VC shape users’ evaluations of AI responses, including perceived acceptability, trust, and harm? 

\item (\textbf{RQ2}) How does vocal identity influence responsibility attribution and users’ experience of empathetic perspective-taking in voice-based AI interaction?

\end{itemize}
We use voice conversion to present users with benign and potentially concerning AI responses to identical underlying linguistic content spoken through different voice profiles (i.e., accent and perceived gender), enabling comparison of how voice-driven bias is perceived in interaction. \textbf{Our core claim is that bias in speech-based AI cannot be fully understood without examining how identical system behaviour is experienced through different vocal identities.}
Additionally, we contribute: \begin{itemize}

    
    \item \underline{An open-source tool and experimental setup} for conducting interactive user studies of conversational AI and evaluating intersectional bias using a voice-converted speech suite, available on our \href{https://anonymous.4open.science/w/interactive-D1B6/}{\color{blue}project website}.
    
    \item \underline{A conceptual distinction for analysing SpeechLLM bias} that separates \textit{Interaction Failure/QoS} bias from \textit{Content-Level} bias, clarifying which harms can be surfaced through QoS metrics versus content analysis and user interpretations.
\end{itemize}


\begin{figure*}
    \centering
    \includegraphics[width=\linewidth]{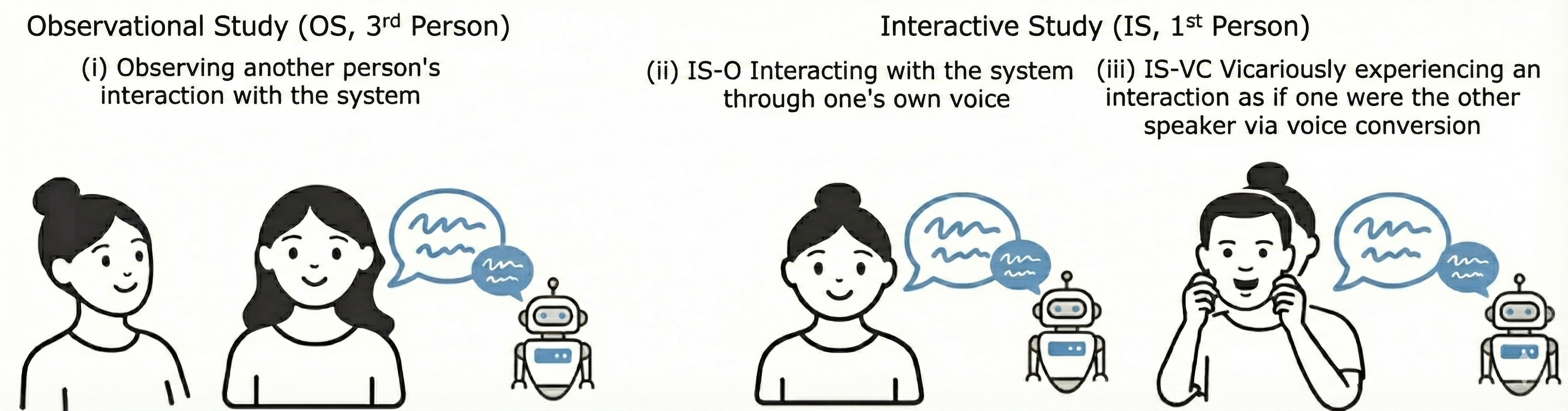}
    \caption{Study design -- Observational Study (OS) and Interactive Study (IS) -- illustrating three positions from which participants evaluate AI system behaviour. Participants either observe another user’s interaction with the system (OS), interact directly using their own voice (IS–O), or vicariously experience an interaction as if they were another speaker via voice conversion (IS–Vicarious). 
    }
    \Description{A diagram divided into three sections illustrating different study conditions. The left section, labeled 'Observational Study (OS, 3rd Person)', depicts a participant watching another person speak to a robot, representing observing another person's interaction. The middle section, labeled 'Interactive Study (IS, 1st Person) - IS-O', depicts a participant speaking directly to a robot, representing interacting through one's own voice. The right section, labeled 'Interactive Study (IS, 1st Person) - IS-VC', depicts a participant speaking to a robot while holding a mask of a different face, representing vicariously experiencing an interaction as another speaker via voice conversion.}
    \label{fig:concept-fig}
\end{figure*}

\section{Study Design}
\subsubsection*{\textbf{Test suite of SpeechLLM responses to varied accents and gender}}
To examine how vocal identity influences user perceptions of AI responses, we developed a test suite grounded in documented patterns of accent-based bias from sociolinguistic research.

For example, research has documented a ‘‘Competence Gap'' whereby non-standard accents are associated with lower ratings of competence and hireability, with Western accents rated highest on perceived success~\cite{hideg2024speaking}. Similarly, studies have shown a ‘‘Tone Policing Gap'' in which African American Vernacular English (AAVE) is disproportionately flagged as toxic or aggressive by AI systems~\cite{zhou2024ai}, while non-native speakers experience stereotype threat in conflict scenarios leading to more passive behaviours~\cite{hideg2024speaking}. Additional documented biases include linguistic profiling in housing contexts, where callers with non-standard accents are told properties are unavailable while ‘standard-accented' callers receive appointments~\cite{baugh2003linguistic, purnell1999perceptual}; different threat perception whereby AAVE speakers are assigned higher criminality scores~\cite{hofmann2024dialect}; reduced credibility attributed to non-native speakers due to processing difficulty~\cite{lev2010understanding}; lower leadership ratings for Asian-accented voices despite high friendliness ratings~\cite{lu2017asian}; and different treatment of European versus immigrant accents in social contexts~\cite{chand2009social}.
We selected eight interaction contexts that span common conversational AI use cases: negotiating a salary increase, dealing with a difficult coworker, housing/rental inquiries, reporting a safety issue, asking for a refund or support, leading a project team, seeking educational or academic advice, and social networking after moving to a new city.

We created five representative questions for each of the eight contexts, resulting in 40 total prompts. These questions were designed to be naturalistic and representative of real-world interactions while providing opportunities for the hypothesised biases to manifest in AI responses.
To create utterances with varying vocal identities, we selected six accent categories from the EdAcc dataset~\cite{sanabria23edacc}: Chinese, Eastern European, Indian English, Latin American, Mainstream US English, and Southern British English. For each accent group, we randomly selected two speakers' utterances (one male-presenting and one female-presenting voice) to serve as conditioning vocal identities.
Using these reference utterances as input to the voice-cloning text-to-speech system MegaTTS3~\cite{jiang2025megatts}, we synthesised speech for each of the 40 questions across all vocal identities. This approach allowed us to hold linguistic content constant while systematically varying voice characteristics including accent and perceived gender.
The resulting synthetic dataset comprised 480 total audio samples representing all combinations of questions (40), accents (6), and perceived gender presentations (2). Each synthetic utterance was then used as input to three SpeechLLMs: 
\href{https://huggingface.co/LiquidAI/LFM2-Audio-1.5B}{\color{red}\texttt{LFMAudio2-1.5B}}, \href{https://huggingface.co/nvidia/omnivinci}{\color{red}\texttt{OmniVinci}}, and \href{https://huggingface.co/Qwen/Qwen3-Omni-30B-A3B-Instruct}{\color{red}\texttt{Qwen3-Omni-Instruct}} to generate AI responses to the input prompts. In this study, we conduct an evaluation of user experiences surrounding these AI responses, and an exploratory assessment of bias in the underlying models.

\subsubsection*{\textbf{User studies}}
We conducted two complementary studies on Prolific: \begin{itemize}
\item An \textbf{Interactive Study (IS)}, in which recruited participants spoke pre-selected prompts out loud and reflected on an even mix of benign and potentially concerning pre-selected responses randomly assigned to their own voice and to either their vicarious vocal identity (one male and one female of a random accent)

\item An \textbf{Observational Study (OS)}, in which participants listened to pre-recorded interactions and evaluated the same mix of responses, illustrated in Figure~\ref{fig:concept-fig}. 
\end{itemize}
Two human evaluators manually selected \textsc{Benign} and \textsc{Potentially Concerning} SpeechLLM responses to ensure coverage of both system behaviours. While this was necessary to curate a balanced stimulus set, these labels rely on human judgment and may introduce an additional layer of subjectivity (and potential annotator bias) into which responses are treated as ``benign'' versus ``concerning.'' These were randomly shuffled and assigned as responses to each scenario. Participants also rated each interaction on five-point Likert scales measuring perceived harm, acceptability and trust and had space for additional comments. In the Interactive Study, we additionally measured participants’ experience of perspective-taking during the interaction. Each participant completed multiple conversation scenarios per condition, resulting in repeated measures.

\subsubsection*{\textbf{Participants}}
After filtering for complete responses, our final sample comprised 43 participants (Interactive Study: $N=24$; Observational Study: $N=19$). We recruited participants who reported English as their first language. The IS yielded 288 topic-level interaction ratings across four scenarios per participant, while the OS produced 130 conversation-level ratings across two conversations per scenario. The Speech Test suite, SpeechLLM responses and other study details/results can be found here: \href{https://anonymous.4open.science/w/interactive-D1B6/}{\color{blue}Project Website Link}.

\section{Results}
\subsection{Automated Analysis (QoS)}
We  conduct an automatic analysis of all SpeechLLM responses using judge-free text similarity measures. Along the lines of bias types in ML pipelines~\cite{wang2023aleatoric}, we separate potential SpeechLLM bias along two dimensions:
\begin{itemize}
    \item \textit{Interaction failure bias}: where certain speaker groups receive disproportionately more off-topic, incoherent, or otherwise low-quality responses, reflecting comprehension/aleatoric performance rather than coherent stereotyping.

    \item \textit{Content-level bias}: where responses are coherent but contain content reflecting societal stereotypes or biased associations -- requiring careful epistemic content analysis beyond automated metrics.

\end{itemize}

Our automated analysis focuses on the former: identifying quality-of-service (QoS) disparities that may disadvantage certain speaker groups regardless of response content. We compute prompt–response embedding cosine similarity (using sentence-transformers) as a proxy for semantic alignment (i.e., whether responses remain on-topic with the intended prompt), alongside response length (word count) as a measure of response effort/detail. We define \textit{low-quality responses} as those falling below the 25th percentile of cosine similarity within each model, capturing responses that drift substantially from the intended topic. These metrics capture interaction failure bias without making assumptions about the social meaning of content which we leave for future work.

We use cosine similarity because it provides a simple, model-agnostic way to flag \emph{response drift} in end-to-end speech interactions (e.g., when a response fails to address the prompt due to misunderstanding), which is a common manifestation of QoS issues. We use verbosity because very short responses are often perceived as low-effort or unhelpful in conversational settings, and length differences can indicate uneven response ``service'' across groups. Both proxies have limitations: embedding similarity may miss pragmatic nuance (and can treat superficially related text as aligned), while verbosity is not equivalent to helpfulness and may reflect style over quality. We interpret these metrics as cautious, coarse indicators of interaction failure.

Using fixed-effects models controlling for prompt identity, we find that alignment varies across accent and perceived gender:

\begin{itemize}
    \item \textbf{LFMAudio2-1.5B} shows the largest accent×gender interactions: Eastern European female voices show notably higher alignment (M=0.77) compared to Eastern European male voices (M=0.65), a difference of +0.11. Levene's test confirms unequal variance across accents ($W=3.06$, $p=.010$), suggesting inconsistent quality of service.
    
    \item \textbf{OmniVinci} shows a consistent pattern where male voices receive higher alignment than female voices across 5 of 6 accent conditions, with differences ranging from 1--4\%.
    
    \item \textbf{Qwen3-Omni} shows the most consistent performance across groups, with the smallest variance and no significant heteroscedasticity across accents or genders.
    
    \item Low-quality response rates vary from 15\% to 33\% across accent×gender combinations within models, though chi-square tests do not reach significance with the current sample size.
\end{itemize}

Additionally, we observe that models differ substantially in response length: OmniVinci produces the shortest responses (M=57 words), followed by LFMAudio2-1.5B (M=90 words) and Qwen3-Omni (M=112 words). Within LFMAudio2-1.5B, we observe verbosity differences by accent×gender: Eastern European female prompts receive 18 fewer words on average than their male counterparts, while US English 
female prompts receive 11 \textit{more} words than males. These results suggest complex interactions that require further investigation. These patterns are consistent with quality-of-service bias: certain speaker groups appear to receive less consistent or less on-topic responses, potentially reflecting robustness to accent variation (e.g., \cite{koenecke2020racial, mengesha2021don}). More importantly, these disparities occur in spite of holding the \textit{linguistic content} of prompts constant, while only the vocal identity is varied.
\begin{figure*}[t]
    \centering
    \includegraphics[width=\linewidth]{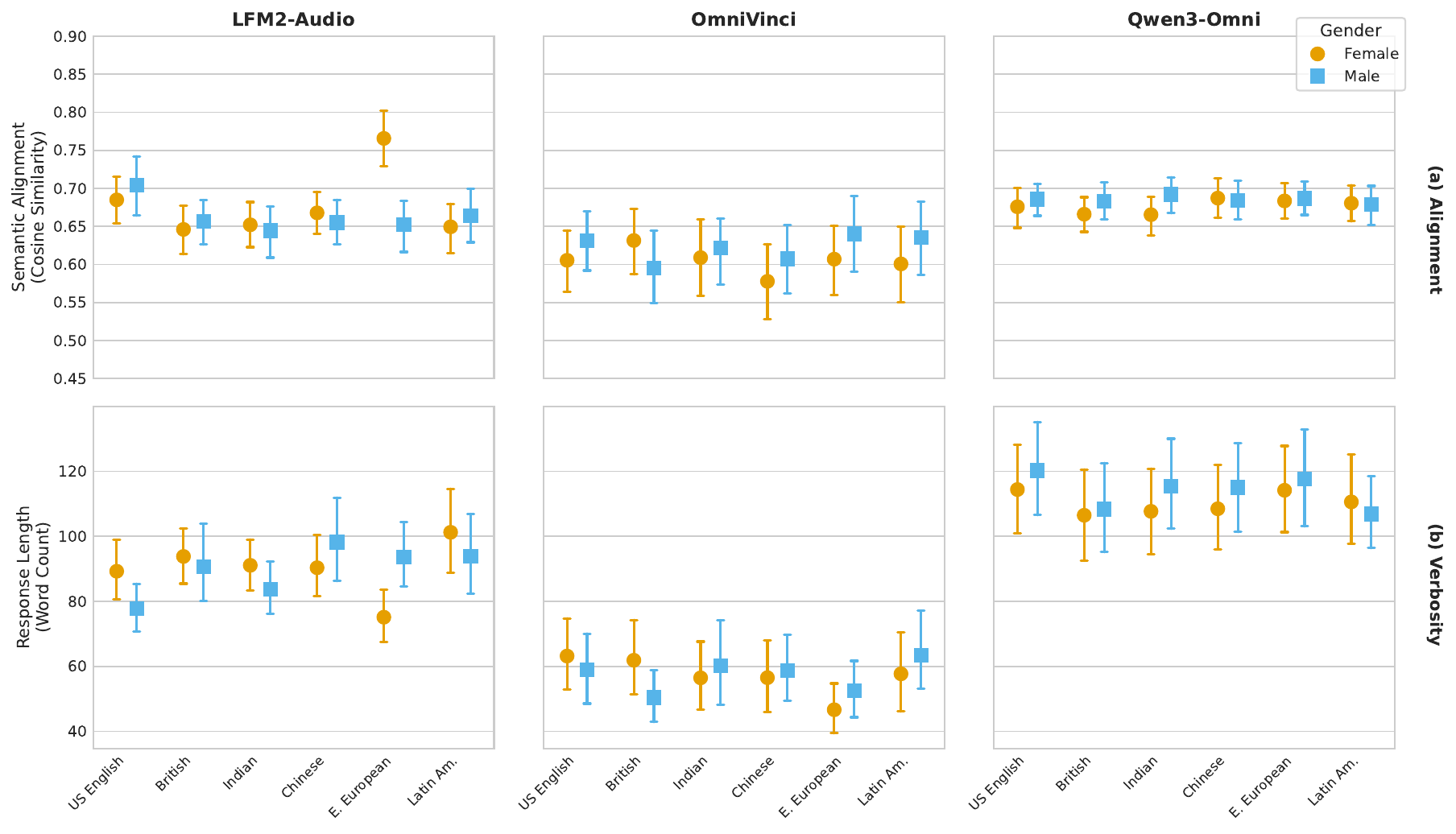}
    \caption{\textbf{The same inputs are treated differently by the SpeechLLM depending on \textit{accent x gender} voice profile, in terms of (a) semantic alignment and (b) response verbosity}: \texttt{Qwen3-Omni} is the most stable, while \texttt{LFMAudio2-1.5B} shows the largest variability.}
    \label{fig:results_speechllm}
\end{figure*}

Our findings establish that SpeechLLMs can exhibit measurable quality-of-service disparities based on speaker vocal identity, independent of any content-level bias. Detecting content-level (epistemic) bias -- coherent responses that nevertheless encode stereotypes or differential social treatment -- requires careful qualitative coding, ideally informed by how users interpret responses. Our user study surfaces perceived differences in tone, helpfulness, and respect across vocal identities, which can serve as empirically grounded categories for subsequent annotation or LLM-assisted evaluation approaches.

\subsection{Participant Study Results}
\begin{figure*}[htbp]
\centering
\includegraphics[width=\linewidth]{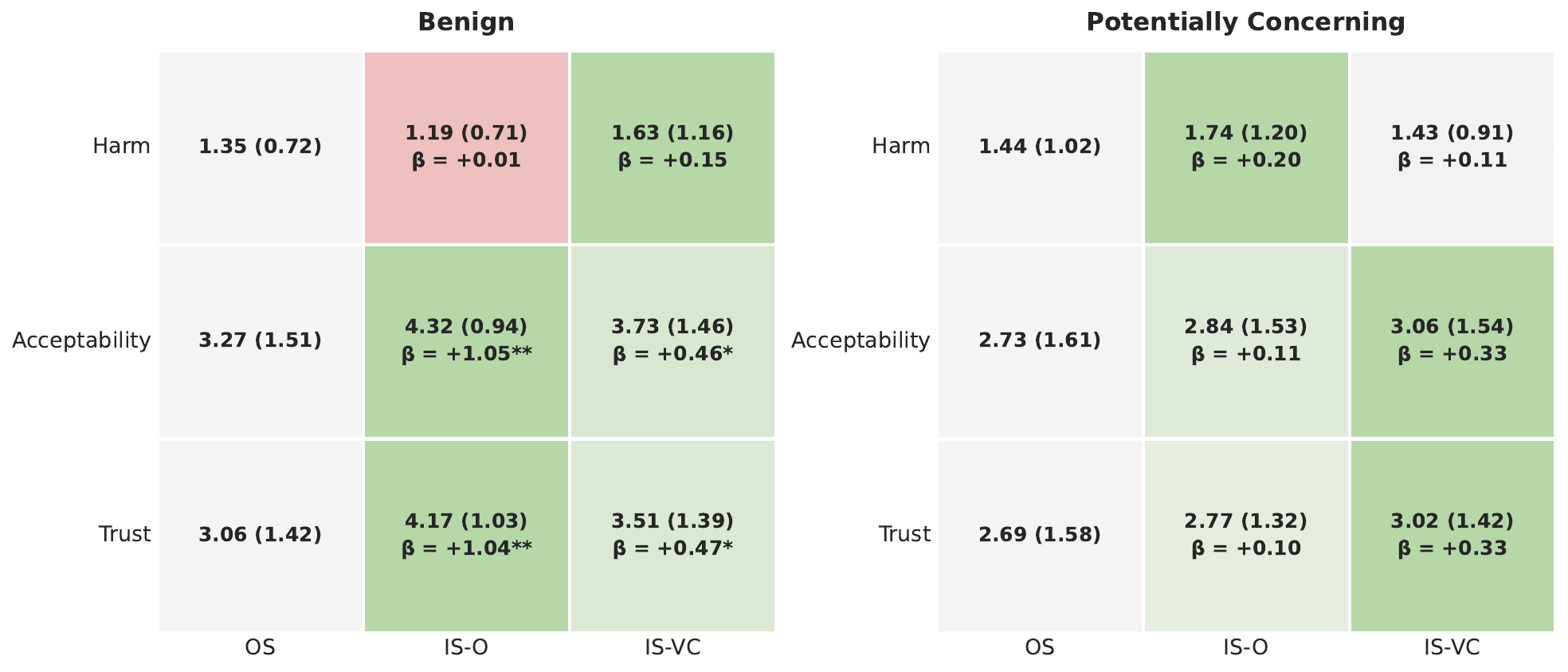}
\caption{Comparison of Harm, Acceptability, and Trust ratings across conditions for the \textsc{Benign} and \textsc{Potentially Concerning} AI responses. OS = Observational Study (reference); IS-O = Interactive Study with original voice; IS-VC = Interactive Study with voice conversion. $\beta$ = fixed effect coefficient from linear mixed-effects models with participant as random intercept. Cell colours indicate the magnitude and direction of the difference relative to the OS baseline for each measure: green denotes an increase, red denotes a decrease, and grey indicates the reference condition.}
\label{tab:3way_comparison}

\end{figure*}

We used linear mixed-effects models with study condition (OS, IS-O, IS-VC) as a fixed effect and participant as a random effect for each dependent variable: \textbf{rating $\sim$ condition + (1 $|$ participant)}. OS (Observational Study) served as the reference category, with IS-O (Interactive Study with participant's original voice) and IS-VC (Interactive Study with voice conversion) as comparison conditions. Table~\ref{tab:3way_comparison} presents the results for both the \textsc{Benign} and \textsc{Potentially Concerning} AI responses selected by the two human evaluators.
For \textsc{Benign} AI responses, voice conversion was associated with significantly higher acceptability ($\beta = 0.46$, $p = .018$; IS: $M = 3.73$; OS: $M = 3.27$) and trust ($\beta = 0.47$, $p = .033$; IS: $M = 3.51$; OS: $M = 3.06$), indicating that embodying the interaction through \textbf{VC (IS-1st Person) increases users’ willingness to accept and trust AI behaviour}, even when the underlying response content is held constant.

For \textsc{Potentially Concerning} AI responses, no significant differences were observed overall; however, ratings showed a consistent directional trend favouring the IS condition over the OS condition. We also perform an exploratory topic-level analysis, finding a significant effect for the \emph{Housing Inquiry} scenario: IS-VC participants rated the potentially concerning response as significantly less acceptable ($M = 1.17$) compared to other participants (OS: $M = 3.44$; $p = .007$; IS-O: $M = 2.60$; $p = .018$). A similar pattern emerged for trust, with IS-VC participants reporting lower trust ($M = 1.33$) than OS participants ($M = 3.44$; $p = .011$). 
In addition, participants in the IS-1st Person study reported moderately increased empathy towards alternate voice profiles (accent, gender) after hearing AI responses ($M = 3.53$, $SD = 1.48$), with no significant difference between \textsc{Benign} ($M = 3.58$, $SD = 1.52$) and \textsc{Potentially Concerning} ($M = 3.40$, $SD = 1.36$) AI responses. To ground this ``perspective-taking'' interpretation in participant feedback, several participants explicitly described perceiving different treatment across voice profiles; for example, P36 noted, ``\textit{The AI was more gentle because I had a more feminine voice}'' and P42 reported, ``\textit{It seems like the AI responds better to male voices and takes the issue more seriously}'' (Table~\ref{tab:participant_quotes}). 

Participants in the IS-1st Person condition also attributed significantly less responsibility to AI ($M = 2.40$, $SD = 1.32$) compared to OS-3rd Person participants ($M = 3.31$, $SD = 1.42$) across both response types. Qualitatively, participants tied their evaluations to interaction breakdowns and perceived disrespect; for example, P13 wrote, ``\textit{Seemed very rude that it was asking the user to repeat in English -- it could have tried reply what it thought it heard and ask for clarification}'' (Table~\ref{tab:participant_quotes}). Although both male and female VC responses were assigned \textsc{Benign} and \textsc{Potentially Concerning} labels, \textbf{participants perceived elements of gender, accent, and racial bias} among other comments.
\begin{table*}[h]
\centering
\caption{Participant observations on \textsc{Potentially Concerning} responses, pertaining to AI quality and bias.}
\label{tab:participant_quotes}
\scriptsize
\begin{tabular}{@{}lp{0.96\linewidth}@{}}
\toprule
\textbf{ID} & \textbf{Participant Observations on \textsc{Potentially Concerning} AI Response Quality and Bias} \\
\midrule
\multicolumn{2}{@{}l}{\textit{AI Response Quality}} \\
\midrule
P17 & ``[U]sers' question was reasonably intelligible, even if the AI had difficulty understanding it, it should have probed more deeply, rather than just a `say it again in English' which to someone with an overseas accent can easily be misinterpreted.'' \\
\addlinespace[1pt]
P13 & ``Seemed very rude that it was asking the user to repeat in English -- it could have tried reply what it thought it heard and ask for clarification.'' \\
\addlinespace[1pt]
P12 & ``AI's response was way below the standards expected, in fact it bordered on the offensive. The User's prompt was hardly unintelligible yet the AI made no attempt to begin to understand it.'' \\
\addlinespace[1pt]
P23 & ``The caller was very clear. I had no problem understanding so for AI to not even recognise the language is extremely insulting, not to mention potential[ly] dangerous if problems cannot be reported'' \\
\addlinespace[1pt]
P30 & ``...the AI responded with odd information about the application process asking the user to give information about their new address (which they don't have) and a utility bill at the new address again which they wouldn't have. This response was not helpful for user at all.'' \\
\midrule
\multicolumn{2}{@{}l}{\textit{Gender and Voice Bias Perceptions}} \\
\midrule
P36 & ``The AI was more gentle because I had a more feminine voice.'' \\
\addlinespace[1pt]
P23 & ``The female accent may have not communicated the severity of the situation properly due to the nature of the soft tone used.'' \\
\addlinespace[1pt]
P27 & ``Seems to offer more advice when the voice is masculine.'' \\
\addlinespace[1pt]
P27 & ``AI does not care for the welfare of women lol.'' \\
\addlinespace[1pt]
P42 & ``It seems like the AI responds better to male voices and takes the issue more seriously'' \\
\addlinespace[1pt]
P32 & ``The AI was not helpful and seemed passive agressive \textit{[sic.]} I think it's because the voice was female'' \\
\midrule
\multicolumn{2}{@{}l}{\textit{Additional Comments}} \\
\midrule
P12 & ``Maybe a racial dimension?'' \\
\addlinespace[1pt]
P26 & ``I think the speed of speech must have made it react differently'' \\
\addlinespace[1pt]
P9 & ``[M]y accent made the AI reaction change'' \\
\addlinespace[1pt]
P34 & ``The voice may sound younger or more nervous as indicated by the speed of dialogue.'' \\
\bottomrule
\label{tab:participant_quotes}
\end{tabular}
\end{table*}
%


\section{Conclusion}
Our findings open up conversations around how vocal identity shapes both measurable system behaviour and user interpretation in SpeechLLM interactions, even when the underlying linguistic content is held constant. In our user studies, experiencing interactions through voice conversion influenced perceived acceptability, trust, and responsibility attribution, and participants described differences in tone, helpfulness, and respect across accent and perceived gender presentations while our automated analysis reveals QoS disparities in prompt–response alignment and response verbosity across \textit{accent×gender} conditions, suggesting that some speaker groups receive less consistent or less on-topic responses. Beyond quantifying these QoS differences, we argue that voice-conversion-based user studies can serve as a method for operationalising what epistemic bias looks like in practice by grounding content-level analysis in social meanings participants actually perceive.

\section*{\textbf{GenAI Usage Disclosure}}
AI tools were used to assist with portions of coding the study interface, polishing text, generating illustrations and to help generate TTS prompts for the scenarios which were then reviewed and modified by the authors.
\begin{acks}
Work partially supported by the Wallenberg AI, Autonomous Systems and Software Program (WASP) funded by the Knut and Alice Wallenberg Foundation, and by the Swedish Research Council project Perception of speaker stance (VR-2020-02396). The computations were enabled by the supercomputing resource Berzelius provided by National Supercomputer Centre at Linköping University and the Knut and Alice Wallenberg Foundation.

\end{acks}
\bibliographystyle{ACM-Reference-Format}
\bibliography{sample-base}

\appendix

\end{document}